\RequirePackage{fix-cm}
\documentclass[smallextended]{svjour3}       
\smartqed  

\usepackage[table]{xcolor}
\usepackage[shadow]{todonotes}
\usepackage[shadow]{rotating}
\usepackage{graphicx}
\usepackage{amsmath}
\usepackage{amssymb}
\usepackage{booktabs}
\usepackage{cite}
\usepackage[section]{placeins}

\newcommand{\R}{\mathbb{R}}
\newcommand{\dx}{\, \mathrm{dx}}

\DeclareMathOperator{\Grad}{\nabla}
\DeclareMathOperator{\tr}{tr}
\newcommand{\rcg}{{\overline{ \bf C} }}
\newcommand{\disp}{{\bf u}}
\newcommand{\mvec}{{\bf m}}
\newcommand{\wvec}{{\bf w}}

\newcommand{\phiv}{{\mathbf\phi}} 

\newcommand{\deltau}{{\bf \delta u}}
\newcommand{\deltaw}{{\bf \delta w}}
\newcommand{\deltam}{{\bf \delta m}}
\newcommand{\deltap}{\delta p}

\begin{document}

\title{Adjoint Multi-Start Based Estimation of Cardiac Hyperelastic Material Parameters using Shear Data}

\titlerunning{Adjoint-based multi-start cardiac parameter estimation}        

\author{Gabriel Balaban \and
        Martin S. Aln{\ae}s \and
        Joakim Sundnes \and
        Marie E. Rognes
}


\institute{G. Balaban, M. S. Aln{\ae}s, J. Sundnes, M. E. Rognes \at
           Simula Research Laboratory \\
           P.O. Box 134 1325 Lysaker, Norway \\
           Tel.: +47 67 82 82 00 \\
           Fax: +47 67 82 82 01 \\
           \email{gabrib@simula.no}           
           \and
           G. Balaban, J. Sundnes \at
           Department of Informatics \\
           University of Oslo \\
           P.O. Box 1080 Blindern 0316 Oslo, Norway
           \and
           M. E. Rognes \at
           Department of Mathematics \\
           University of Oslo \\
           P.O. Box 1053 Blindern 0316 Oslo, Norway
}

\date{}

\maketitle

\begin{abstract}
Cardiac muscle tissue during relaxation is commonly modelled as a
hyperelastic material with strongly nonlinear and anisotropic stress
response.  Adapting the behavior of such a model to experimental or
patient data gives rise to a parameter estimation problem which
involves a significant number of parameters. Gradient-based
optimization algorithms provide a way to solve such nonlinear
parameter estimation problems with relatively few iterations, but
require the gradient of the objective functional with respect to the model
parameters. This gradient has traditionally been obtained using finite
differences, the calculation of which scales linearly with the number
of model parameters, and introduces a differencing error. By using an
automatically derived adjoint equation,
we are able to calculate this
gradient more efficiently, and with minimal implementation effort. We
test this adjoint framework on a least squares fitting problem
involving data from simple shear tests on cardiac tissue samples.
A second challenge which
arises in gradient-based optimization is the dependency of the
algorithm on a suitable initial guess. We show how a multi-start
procedure can alleviate this dependency. Finally, we provide estimates
for the material parameters of the Holzapfel and Ogden strain energy
law using finite element models together with experimental shear data.
\end{abstract}



\section{Introduction}
\label{intro}

The personalization of computational models in cardiology is a key
step towards making models useful in clinical practice and cardiac surgery.
A computational model, once properly calibrated, has the potential to forecast cardiac
function and disease, and can aid in planning treatments and
therapies.  To describe the mechanical function of the heart, the
passive elasticity of the muscle tissue needs to be
represented. Personalizing the effects of this elasticity in a
computational model is typically accomplished by tuning a set of
material parameters so that the output of the model fits observed
data.  Gradient-based optimization algorithms have successfully been
used in the past to automatically perform the parameter tuning at an
organ scale~\cite{augenstein2005method, Wang2009}. In these studies,
the gradient of the objective functional is approximated using
one-sided finite differences.

Compared to using a global optimization method, local gradient-based
methods have the advantage of using relatively few optimization
iterations. This is an important consideration when optimizing organ
scale finite element models, for which running a single forward model
can take hours or days. On the other hand, a disadvantage of using
local optimization methods is the fact that they can converge to
local, globally suboptimal, minima. One way to combine the speed of a
local optimization with the robustness of a global optimization is to
use the multi-start method. In this method, many local optimizations
are run starting from various points in parameter space and the best
fitting solution of the group is taken to be the global optimum.

Another popular approach to parameter fitting is the reduced order
unscented Kalman filter. This approach was successfully used to fit a
transversely isotropic passive mechanics model to synthetic
data~\cite{xi2011myocardial}, to partially calibrate a multi-physics
model~\cite{Marchesseau2013}, and to estimate regional contractility
parameters~\cite{chabiniok2012estimation}. Note however that the use
of both unscented Kalman filtering and finite differences carries a
computational cost that increases with the number of model parameters.

Assuming there are $k$ parameters to be estimated, an
unscented Kalman filter with a minimal sigma-point configuration
requires $k + 1$ model evaluations at a single time level for each
assimilated data point. An evaluation of a finite difference
derivative on the other hand requires $k + 1$ runs of the model
throughout the full span of model configurations considered.

In contrast to these two techniques, the adjoint approach computes the
objective functional gradient via the solution to an adjoint equation,
which involves only a single solve of a linearized system for
any number of model parameters. Thus, for models involving many
parameters, either due to model complexity or spatiotemporal parameter
variation, the adjoint approach offers a computationally attractive
approach for parameter estimation.

There are some previous results involving adjoint equations and
cardiac elasticity. Sundar et al. (2009) developed a framework for the
estimation of wall motion based on cine-MRI images and adjoint
inversion~\cite{sundar2009biomechanically}, and Delingette et
al. (2012) used an adjoint equation to estimate contractility
parameters~\cite{Delingette2012}.  However, both of these studies
involve linear and isotropic elasticity models, which represent a
significant simplification of the orthotropic and highly nonlinear
behavior reported in the contemporary cardiac mechanics
literature~\cite{costa2001modelling, dokos2002shear, Holzapfel2009}.

One reason why it is difficult to use an adjoint equation with modern
nonlinear anisotropic models is the complexity required in deriving
and implementing code for the solution of the adjoint problem. In
order to resolve this issue, we make use of an automatic framework for
generating adjoint code~\cite{farrell2013automated}. Here, we use this
adjoint framework to estimate the material parameters of an
invariant-based orthotropic myocardial strain energy law (the
Holzapfel-Ogden model)~\cite{Holzapfel2009}. This law is embedded here
in an incompressible finite element framework, and we use the raw data
from a simple shearing experiment~\cite{dokos2002shear} as a target
for optimization. These data have previously been used to estimate
material parameters for a variety of other strain energy functions
using a finite element framework, but with a gradient obtained using
finite differences~\cite{schmid2008myocardial,
  schmid2009myocardial}. The material parameters of the particular
strain energy density that we are using have also been previously
estimated using digitized data based on Figure 6
of~\cite{dokos2002shear}, and a homogeneous deformation
model~\cite{Holzapfel2009,wang2013structure,goktepe2011computational}. Our
study is however the first to use the adjoint approach for the
estimation of cardiac hyperelasticity parameters and the first to
provide optimized material parameters for the incompressible
Holzapfel-Ogden model for non-homogeneous shear deformations.

The rest of this paper is organized as follows. In
Section~\ref{sec:methods} we describe the variational formulation of
the elasticity model, the optimization problem for identifying the
material parameters, and how the adjoint gradient formula can be used
to calculate a functional gradient. In Section~\ref{sec:results} we
describe the verification of the forward and inverse solvers, present
timings to show the efficiency of the adjoint method, and show the
results of parameter estimations. Finally, we test a multi-start
optimization method in order to reduce the dependence of the
gradient-based algorithm on the choice of initial parameter set. We
conclude by discussing our findings in Section~\ref{sec:discussion},
and drawing some conclusions in Section~\ref{sec:conclusion}.


\section{Mathematical models and methods}
\label{sec:methods}
We shall use the notion of the directional derivative frequently
throughout. For a functional $f : Y \rightarrow \R$ for some vector
space $Y$, we define the directional derivative of $f$ with respect to
the argument named ${\bf y}$ in the direction $\delta {\bf y}$
\begin{equation*}
  D_{\bf y} f ({\bf y}) [{\bf \delta y}]
  \equiv \frac{\partial}{\partial \epsilon} f({\bf y} + \epsilon \ {\bf \delta y})\Big|_{\epsilon = 0}.
\end{equation*}
Furthermore we denote the total derivative by the usual notation
$ \frac{Df}{Dy}$
to mean the derivative of $f$ with respect to all arguments depending on $y$.


\subsection{Hyperelasticity model}
Let $\Omega \subset \R^3$ be an open and bounded domain with
coordinates ${\bf X}$ and boundary $\partial \Omega$, occupied by an
incompressible hyperelastic body. We consider the quasi-static regime
of a body undergoing a large deformation ${\bf x} = {\bf x}({\bf X})$ and are interested in finding
the displacement $\disp = \disp({\bf X}) = {\bf x} - {\bf X}$ and the hydrostatic pressure
$p = p({\bf X})$ that minimize the incompressible strain energy $\Pi =
\Pi(\disp, p, \mvec)$:
\begin{equation}
  \label{eq:energy}
  \Pi (\disp, p, \mvec) = \int_{\Omega} \psi( \rcg, \mvec) +  p (J - 1) \dx
\end{equation}
over the space of admissible displacements and pressures satisfying
any given Dirichlet boundary conditions. In~\eqref{eq:energy}, $\mvec$ is a set of material
parameters, $J = \det {\bf F}$, where ${\bf F} = \Grad {\bf x} = \Grad \disp + {\bf I}$ denotes the
deformation gradient, ${\bf I}$ is the identity tensor in $\R^3$,
$\rcg = J^{- \frac{2}{3}} {\bf F}^{T} {\bf F}$ denotes a
volume-preserving right Cauchy-Green strain tensor, and $\psi$ denotes an
isochoric strain energy density.

The incompressible Holzapfel and Ogden hyperelasticity
model~\cite{Holzapfel2009} describes large deformations and stresses
in cardiac tissue via the following energy density $\psi$:
\begin{equation}
  \begin{aligned}
    \label{eq:hao}
    \psi(\overline{\bf{C}}, \mvec) = &\, \frac{a}{2 b} \left( \exp \left[ b (I_1( \overline{\bf{C}}) - 3) \right]  -1 \right) \\
    &+ \sum_{i = f,s }\frac{h(I_{4i}(\overline{\bf{C}})) a_i}{2 b_i} \left(\exp \left[ b_i (I_{4i}(\overline{\bf{C}}) - 1)^2 \right]
    - 1 \right) \\
    &+ \frac{a_{fs}}{2 b_{fs}} \left(\exp \left[b_{fs} I^2_{8fs}(\overline{\bf{C}}) \right] - 1 \right).
  \end{aligned}
\end{equation}
Here ${f,s}$ denote fiber and sheet directions, respectively; $h(x)$
is a Heaviside function with a jump at $x = 1$, and the material
parameters are
\begin{equation}
\mvec = (a, b, a_f, b_f, a_s, b_s, a_{fs}, b_{fs}).
\end{equation}
Moreover, $I_1, I_{4s}, I_{4f}, I_{8fs}^2$ are rotation invariant
functions given by
\begin{equation}
  \begin{aligned}
    I_1(\rcg) &= \tr \rcg \\
    I_{4i}(\rcg) &= {\bf e}_i \cdotp \rcg {\bf e}_i \quad i = f,s \\
    I_{8fs}(\rcg) &= {\bf e}_s \cdotp \rcg {\bf e}_f
  \end{aligned}
\end{equation}
where $\tr$ denotes the tensor trace and ${\bf e}_f, {\bf e}_s$ denote
unit vectors pointing in the local myocardial fiber and sheet
directions~\cite{Holzapfel2009}.  The strain energy density $\psi$ is
rotation-invariant, and polyconvex if $\mvec >
\mathbf{0}$~\cite{Holzapfel2009}.

The Euler-Lagrange equations for the minimizing displacement $\disp$
and pressure $p$ of~\eqref{eq:energy} read: for given $\mvec$, find
$\wvec = (\disp, p)$ such that
\begin{equation}
  \label{eq:varform_fully}
  R(\wvec, \mvec; \deltaw) \equiv  D_{\disp, p} \Pi (\disp, p, \mvec)[\deltau, \deltap] = 0,
\end{equation}
for all admissible virtual variations $\deltaw = (\deltau,
\deltap)$. Inserting the total potential energy from \eqref{eq:energy}
and taking the directional derivatives, we obtain

\begin{equation}
\label{eq:weakform}
D_{\disp, p} \Pi (\disp, p, \mvec)[\deltau, \deltap]
  = \int_{\Omega} \left(\left(\frac{\partial \psi(\rcg, \mvec)}{\partial {\bf F}} + p J {\bf F}^{-T} \right) : \Grad \deltau
    + (J - 1) \deltap\right) \dx .
\end{equation}

\subsection{Parameter estimation as a PDE-constrained optimization problem}
\label{subsec:opt}
In the general case, the passive material parameters $\mvec$
entering the constitutive relationship~\eqref{eq:hao} are not
known. In order to estimate these parameters from data, we propose to
use a numerical approximation in combination with a gradient-based
optimization algorithm in which the gradients are computed via an
adjoint model. The optimization algorithm seeks to minimize the misfit
between model output and observations. Denoting the misfit functional
by $I = I(\wvec(\mvec), \mvec)$, the optimization problem reads:
\begin{equation}
  \label{eq:optprob}
  \min_\mvec I(\wvec(\mvec), \mvec) \qquad \text{subject to} \qquad
   R(\wvec, \mvec; \deltaw) = 0 \quad \forall \deltaw \in W,
\end{equation}
together with suitable Dirichlet boundary conditions on $\wvec$. We
also require that $\mvec > 0$ to ensure the
functional~\eqref{eq:energy} is polyconvex~\cite{Holzapfel2009}.  For
notational convenience we will sometimes use the reduced formulation
of the misfit functional and its gradient with respect to the material
parameters $\mvec$. In particular, we introduce the reduced functional
$\hat{I}$
\begin{equation}
  \label{eq:Igrad}
    \hat{I}(\mvec) \equiv I(\wvec(\mvec), \mvec). \\
\end{equation}
In our numerical experiments we use Sequential Least Squares
  Programming (SLSQP) as implemented in~\cite{KraftSQP} and wrapped
in the package SciPy~\cite{ScipyPackage} in order to
solve~\eqref{eq:optprob}.


\subsection{Multi-start Optimization}
\label{sec:multistart}
A common challenge with gradient-based algorithms is that the solution
obtained depends on the choice of initialization point for the
algorithm. Moreover, the optimized solution may be a local minimum
only and not necessarily a global minimum.  One way to attack these
issues is to run many optimizations from randomly chosen initial
parameter points, and to chose the resulting optimized material
parameter set that gives the best fit. This method is often referred
to as multi-start optimization~\cite{boender1987bayesian} and is an
example of combining global and local optimization.

Due to the presence of exponential functions in the strain
energy~\eqref{eq:hao}, it is possible for calculated stresses to
become very large, which may result in convergence issues for the
numerical solution of the Euler-Lagrange
equation~\eqref{eq:varform_fully}. This can easily occur if several
material parameters have large values. In order to minimize this
problem we have designed a procedure to generate random initial
guesses which limits the number of large material parameter values
while still allowing for a large range of initial possible values for
each parameter. The procedure works as follows: first set a maximum
parameter value $P_{max}$. Then choose $N$ (with $N = 8$ in our case)
points $p_i, \ i \in \{1,2,3...n \}$, from a uniform distribution defined over the interval $[0,
  P_{max}]$ and let $p_0 = 0$. The parameter values $m_i$ are then set to be the distances
between successive randomly drawn points, that is $m_i = p_i - p_{i - 1}$.

\subsection{Computing the functional gradient via the adjoint solution}

Gradient-based optimization algorithms in general, and the SLSQP
algorithm in particular, rely on the total derivative of the objective
functional~\eqref{eq:Igrad}. By introducing an \emph{adjoint} state
variable, this derivative may be computed efficiently. We
summarize this result below. Our presentation is based
on~\cite{gunzburger2003perspectives}, and is adapted here to the solid
mechanics setting.

We define three abstract spaces $W$, $M$, and $\Phi$, where $W$ is the
space of all possible solutions to the variational
equation~\eqref{eq:varform_fully} which also satisfy any given
Dirichlet boundary conditions, $M$ is the material parameter vector
space, and $\Phi$ is the space of virtual variations.  The Lagrangian
$L: W \times M \times \Phi \rightarrow \R$ is defined as:
\begin{equation}
  \label{eq:Ldef}
  L(\wvec, \mvec, \phiv) = I(\wvec, \mvec) - R(\wvec, \mvec; \phiv).
\end{equation}
For all $\mvec\in M$, $\wvec\in W$ solving the state
equation~\eqref{eq:varform_fully}, we have
\begin{equation*}
  \frac{D}{D \mvec} R(\wvec(\mvec), \mvec; \phi) = 0,
\end{equation*}
such that the total derivatives of $I$ and $L$ coincide,
\begin{equation}
  \label{eq:DIeqDL}
 \frac{D}{D \mvec} I(\wvec(\mvec), \mvec) = \frac{D}{D \mvec} L(\wvec(\mvec), \mvec, \phiv).
\end{equation}
If we choose $\phiv \in \Phi$ such that
\begin{equation}
  \label{eq:adjointdef}
  D_{\wvec} L(\wvec, \mvec, \phiv)[\delta \wvec] = 0
\end{equation}
for all $\delta \wvec \in W$, which in particular includes $\delta
\wvec = D_{\mvec} \wvec(\mvec)[\deltam]$, the total derivative of $L$ with
respect to $\mvec$ in the direction $\deltam$ simplifies as follows
using the chain rule:
\begin{equation}
  \label{eq:DLchainrule}
  \begin{split}
    \frac{D}{D \mvec} L(\wvec(\mvec), \mvec, \phiv)
    &= D_{\wvec} L(\wvec, \mvec, \phiv) [ D_{\mvec} \wvec(\mvec)[\deltam] ]
    + D_{\mvec} L(\wvec, \mvec, \phiv)[\deltam] \\
    &= D_{\mvec} L(\wvec, \mvec, \phiv)[\deltam]
  \end{split}
\end{equation}
Then, for any infinitesimal variation in the material parameters
$\deltam$, combining~\eqref{eq:DIeqDL},~\eqref{eq:DLchainrule},
and~\eqref{eq:Ldef} yields an efficient evaluation formula, not
requiring derivatives of the state variable $\wvec$ with respect to
the material parameters $\mvec$, for the total derivative of $I$:
\begin{equation}
  \label{eq:adj_grad}
  \frac{D}{D \mvec} I(\wvec(\mvec), \mvec)
  = D_{\mvec} I(\wvec, \mvec)[\deltam] - D_{\mvec} R(\wvec, \mvec, \phiv)[\deltam].
\end{equation}
We still need to compute $\phiv$. By defining the form $R_{\wvec}$
and its adjoint $R_{\wvec}^*$,
\begin{equation*}
  \begin{split}
R_{\wvec} (\wvec, \mvec; \delta \wvec, \phiv) &\equiv D_{\wvec} R(\wvec, \mvec; \phiv)[\delta \wvec], \\
R_{\wvec}^{*} (\wvec, \mvec; \phiv, \delta \wvec) &\equiv R_{\wvec} (\wvec, \mvec; \delta \wvec)[\phiv],
  \end{split}
\end{equation*}
we can rewrite~\eqref{eq:adjointdef} as
\begin{equation*}
  \begin{split}
  D_{\wvec} L(\wvec, \mvec, \phiv) [\delta \wvec]
  &= D_{\wvec} I(\wvec, \mvec)[\delta \wvec] - R_{\wvec}^{*} (\wvec, \mvec; \phiv, \delta \wvec) = 0,
  \end{split}
\end{equation*}
and thus recognize the adjoint equation: given $\mvec$, $\wvec$, find
$\phiv \in \Phi$ such that
\begin{equation}
  \label{eq:adjoint}
  R_{\wvec}^{*} (\wvec, \mvec; \phiv, \delta \wvec) = D_{\wvec} I(\wvec, \mvec)[\delta \wvec]
\end{equation}
for all $\delta \wvec \in W$.

In summary, the adjoint-based gradient evaluation formula is: given
$\mvec$, first compute $\wvec$ by solving the state
equation~\eqref{eq:varform_fully}, next compute $\phiv$ by
solving~\eqref{eq:adjoint}, and finally evaluate~\eqref{eq:adj_grad}.

\subsection{Description of shearing experiments}

We aim to optimize the material parameters of the Holzapfel-Ogden
model~\eqref{eq:hao} with respect to target experimental data, in
particular data resulting from an earlier set of simple shearing
experiments~\cite{dokos2002shear}. In these experiments, $6$ pig
hearts were extracted. From each heart, three adjacent $3 \textrm{mm}
\times 3 \textrm{mm} \times 3 \textrm{mm}$ cubic blocks were cut in
such a way that the sides of the cubes were aligned with the local
myocardial fiber and sheet directions.  A device held two opposing
faces of each cube between two plates using an adhesive.  The top
plate was displaced in order to put each specimen in simple shear.
For each specimen 6 different modes of shear were tested. These modes
are described using the $F,S,N$ coordinate system, which refer to the
myocardial fiber, sheet and sheet normal directions,
respectively. Each mode is denoted by two letters, where the first
defines the normal of the face of the cube that is being displaced,
and the second refers to the direction of displacement. These 6 modes
are $FS, FN, SF, SN, NF, NS$.

In order to remove the effects of strain softening, preliminary
displacements were applied to the tissue samples until no further
softening was observed.  After that, displacements were once again
applied, and the forces in the shear direction were measured on the top
plate. These measurements were taken for circa $200-250$ various
states of shear per mode.

\begin{figure}
   \includegraphics[width=1.0\textwidth]{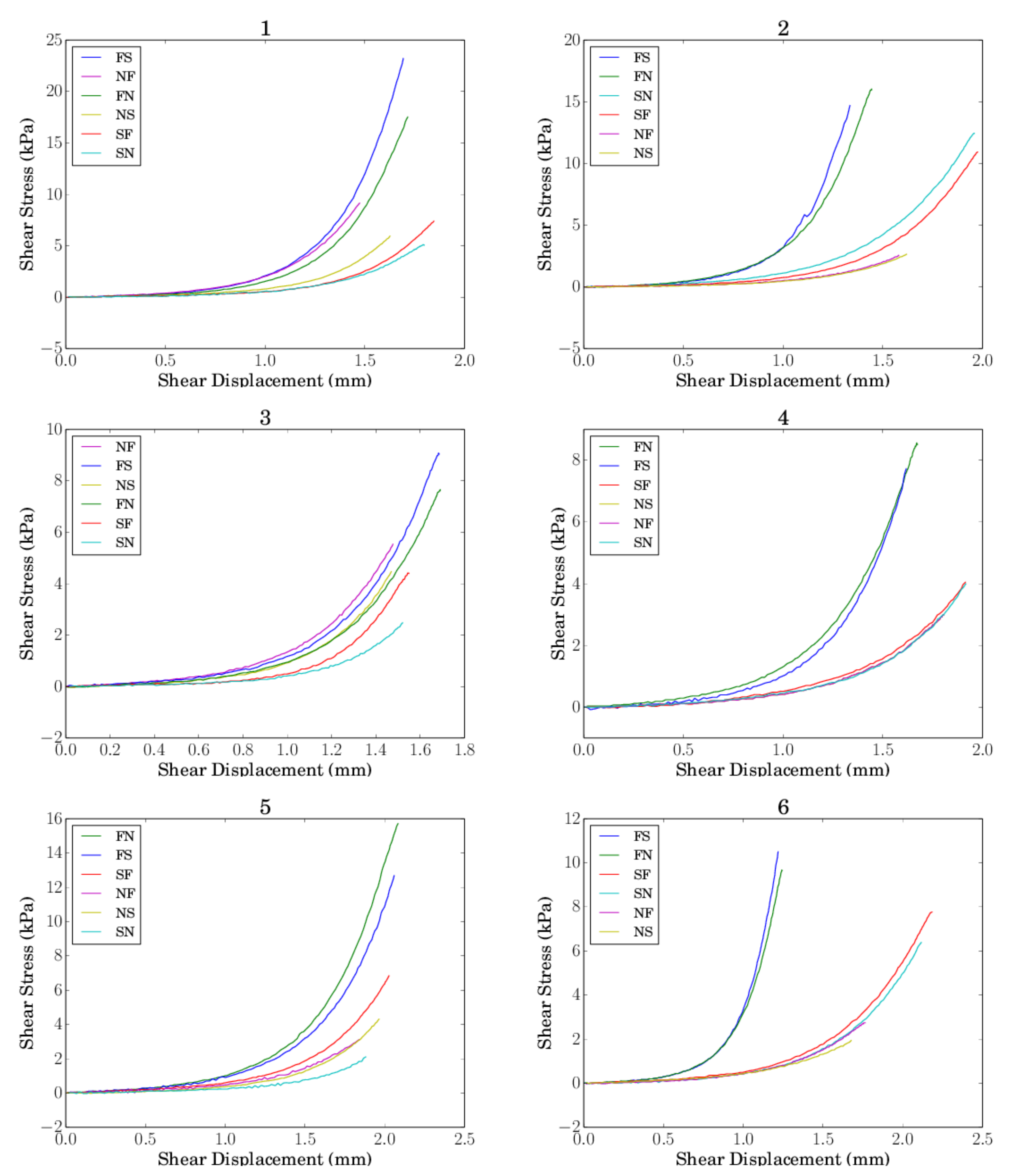}
   \caption{Stress-strain relations, numbered $1$ through $6$, obtained
	    from simple shearing experiments performed on $3 \textrm{mm}
	    \times 3 \textrm{mm} \times 3 \textrm{mm}$ cubes of myocardium
	    extracted from 6 porcine hearts. The modes are ordered from
	    highest to lowest stiffness in each experiment. The data
	    originates from the study~\cite{dokos2002shear}, but were not
	    published in the subsequent article. In Experiment 4 the data for
	    one of the NS-NF curves was copied into the other before we
	    received it, so the two curves lie here on top of one another.}
   \label{fig:shear_exprdata}
\end{figure}

In Figure~\ref{fig:shear_exprdata} we display the stress-strain
relations for positive displacements that were obtained from the
shearing experiments~\cite{dokos2002shear}. As can be seen in
  Figures 4 and 6 of~\cite{dokos2002shear} the experimentally
obtained curves contain a high degree of symmetry through the line $y
= -x$.  We can expect the same symmetry in the stresses computed by
finite element models which use the strain energy~\eqref{eq:hao} since
changing the sign of the displacement map will change the sign of the
resulting stresses but preserve their magnitude. In the previous
studies \cite{Holzapfel2009}, \cite{goktepe2011computational}, and
\cite{wang2013structure}, only the data for positive shear
displacements were used. For the sake of comparability, we restrict
our data in the same way.

In our numerical experiments we use two data sets with reference to
the numbering of~\cite{dokos2002shear}. The first is Data Set 6, and
the second data is Data Set 2 with the $SF$ and $SN$ curves
swapped. This swap and the choice of data sets are discussed further
in Section~\ref{sec:discussion}. For clarity, we shall refer to Data
Set 6 as "transversely isotropic" and Data Set 2 with the swap as
"orthotropic", as the respective stress-strain curves are typical of
materials of these types. For each mode, the prescribed shear
displacement is modelled as a Dirichlet boundary condition for the
displacement on the respective top and bottom faces in the respective
direction.

\subsection{Choice of objective functional}
 
In order to estimate the passive material parameters of the
Holzapfel-Ogden model, we make use of a least squares objective
functional. This functional defines a distance from the model output
to the data points of the shearing experiment, and we seek the
material parameter set $\mvec$ that minimizes this. Before introducing
our objective functional, we define the set of directions $\mathcal{D}
= \{F, S, N\}$, referring to fiber, sheet and sheet normal
directions. We also use the notation $(i,j)$ to refer to a mode, with
the index $i$ referring to the normal of the face that is shifted, and
$j$ to the direction in which the shift occurs.

Our fit function is similar to that used
in~\cite{schmid2007computationally}, and is given by
\begin{equation}
  \label{eq:gaussfit}
  \hat{I}(\mvec)^2 = \sum_{i \in \mathcal{D}} \sum_{j \in \mathcal{D}} \sum_{k=1}^G
  \omega_k \left( t^{i,j}_{\textrm{model}}(c_k, \mvec) - t^{i,j}_{\textrm{exper}}(c_k) \right)^2
\end{equation}
In~\eqref{eq:gaussfit}, $t^{i,j}_{\textrm{exper}}$ is the force
measured during the experiment, and $t^{i,j}_{\textrm{model}}$ is the
force generated by the finite element model at each prescribed shear
displacement $c_k \in [0, C^{i,j}]$, where $C^{i,j}$
is the maximal prescribed displacement of the mode $(i,j)$ in the
experiment. Each $c_k$ is chosen to be a Gauss point of a
$G$-point Gauss integration rule defined over $[0,
  C^{i,j}]$, and $\omega_k$ is the value of the Gauss weight
related to $c_k$. Explicitly, for mode $(i, j)$ with top face
$\partial \Omega_i$, $t^{i,j}_{\textrm{model}}$ is given by
\begin{equation}
  t^{i,j}_{\textrm{model}}(c_k, \mvec) = \int_{\partial \Omega_{i}}
  \frac{\partial \psi (\disp(c_k), \mvec)}{\partial \mathbf{F}_{i,j}} \textrm{\, dS},
\end{equation}
where $\mathbf{F}_{i,j} = \mathbf{e}_{i} \cdotp \mathbf{F}
\mathbf{e}_{j}$ is a shear component of the deformation gradient.
Evaluating the inner loop of $\hat{I}$ requires
solving~\eqref{eq:varform_fully} once for each given shear
displacement $c_k$. The motion given by the calculated displacements
is then a quasi-static approximation of the motion undergone by the
corresponding tissue in the shearing experiment.

Following~\cite{schmid2007computationally}, we evaluate the least
squares fit~\eqref{eq:gaussfit} at $G$ Gauss integration
points, rather than for all 250 recorded points for each shear
mode, in order to greatly reduce the computational expense of evaluating $\hat{I}$.
At each Gauss point we obtain the corresponding shear stress by
linearly interpolating between the two neighbouring stresses which
were recorded in the experiments of Dokos et al.~\cite{dokos2002shear}.

The use of Gauss integration is based on the observation
that $\hat{I}(\mvec)$ is an approximation to the following expression
\begin{equation}
\left( \sum_{j \in \mathcal{D}} \sum_{i \in \mathcal{D}} \int_0^{C^{i,j}}
\left( t^{i,j}_{\mbox{model}}(c, \mvec) - t^{i,j}_{\mbox{exper}}(c) \right)^2 \ dc \right)^{\frac{1}{2}}.
\label{eq:intfit}
\end{equation}


By setting $t^{i,j}_{\mbox{model}} = 0$ and approximating the integral by the
midpoint rule applied to the full dataset we can determine the quality
of the Gauss approximation.  In order to do this we define the
relative error
\begin{equation}
 \epsilon_{rel} = \left| \frac{ \hat{I} - \hat{I}_{mid}}{\hat{I}_{mid}}  \right|,
\end{equation}
where $I_{mid}$ is the midpoint rule approximation
  of~\eqref{eq:intfit} evaluated over the full data, and $I$, given
  by~\eqref{eq:gaussfit}, is evaluated at a reduced set of Gauss
  points. We noticed that 9 Gauss points are sufficient to reduce
  $\epsilon_{rel}$ to less than $0.01$. However, in our numerical
  experiments we use $G = 40$ Gauss points as this guaranteed small
  enough changes in the solution of the Euler-Lagrange
  equation~\eqref{eq:varform_fully} from one Gauss point to the next,
  so that our Newton's method solution of~\eqref{eq:varform_fully}
  always converged.

\subsection{Finite element discretization of the hyperelasticity equations}

We represent each tissue sample of the shearing experiments by a
three-dimensional cube $\Omega = [0, 3]^3$ ($\textrm{mm}^3$).  An $N
\times N \times N$ mesh of this cube was constructed by uniformly
dividing the mesh into $N \times N \times N$ boxes and then
subdividing the boxes into tetrahedra. The local myocardial fiber
  and sheet orientations were represented as spatially constant
  vectors aligned with the coordinate axes.

On these geometries, we solve~\eqref{eq:varform_fully} and its
adjoint, using a Galerkin finite element method with the Taylor-Hood
finite element pair~\cite{hood1974navier}; e.g.~a continuous piecewise
quadratic vector field for the displacement and a continuous piecewise
linear scalar field for the pressure. For the solution of the
nonlinear system of equations, we use a Newton trust region
method. The absolute tolerance of the nonlinear solver was set to
$10^{-10}$ in the numerical experiments below.  Linear systems are
solved by LU factorization.

\begin{center}
\begin{figure}
  \includegraphics[width=0.49\textwidth]{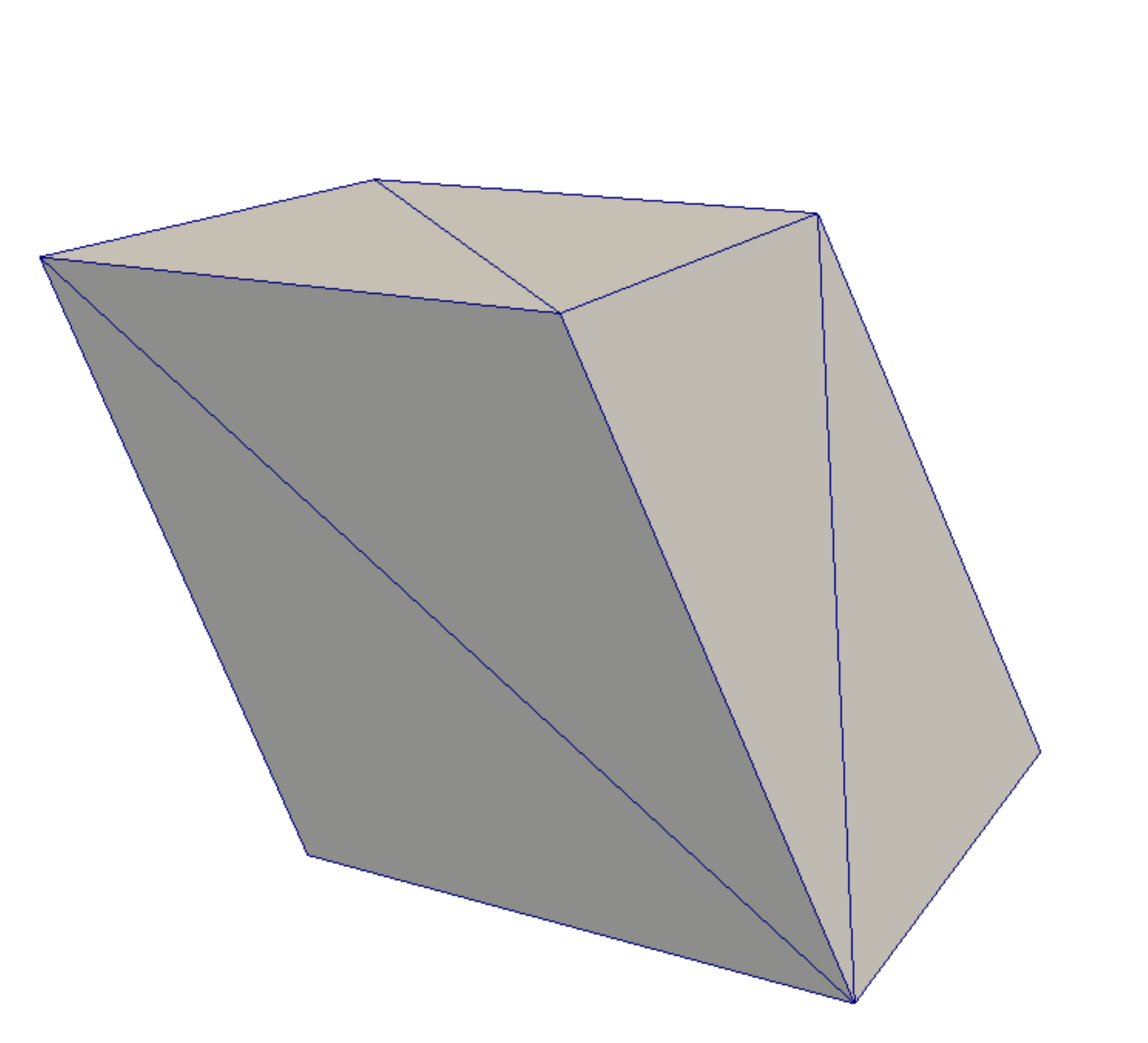}
  \includegraphics[width=0.49\textwidth]{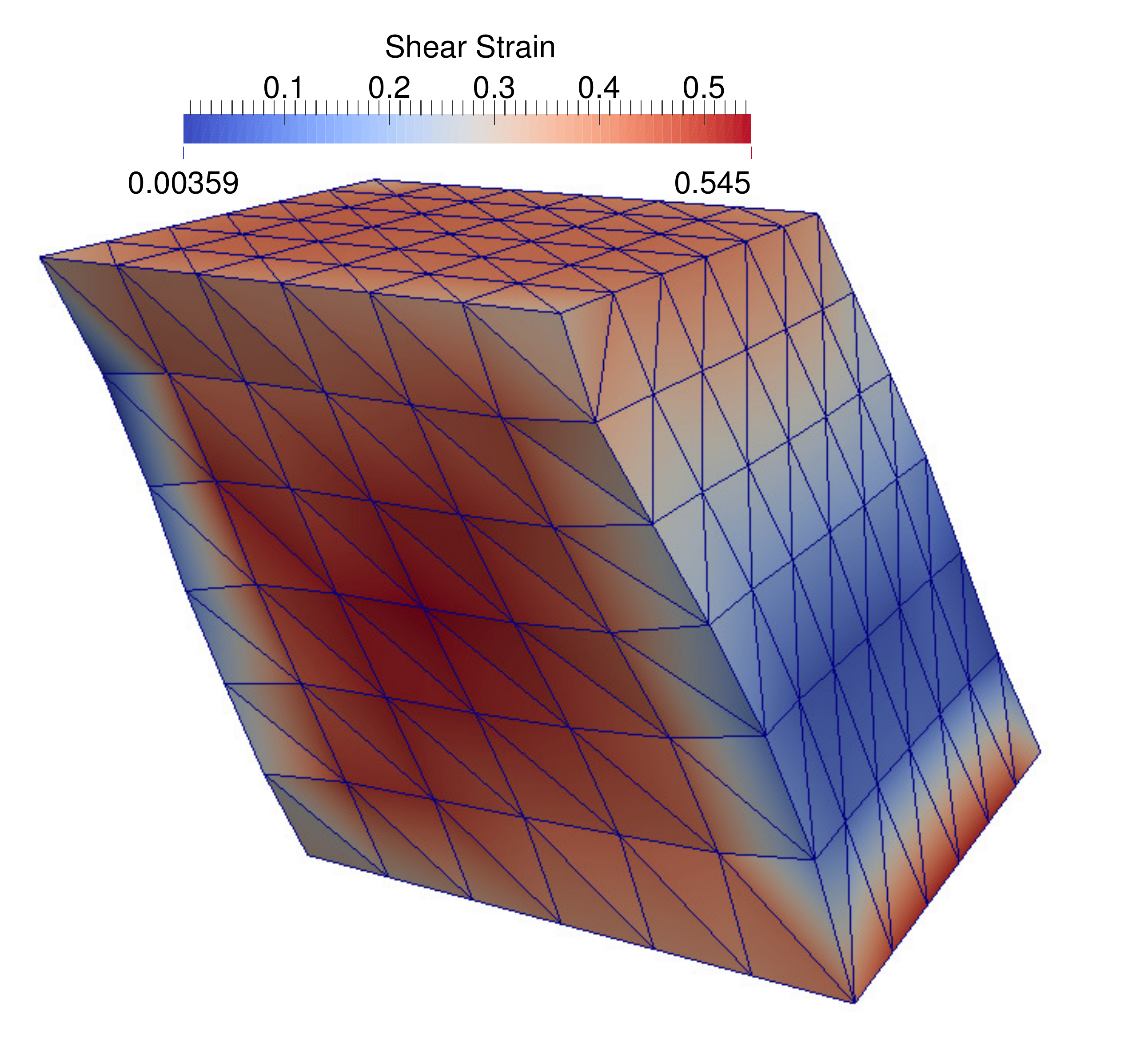}
  \caption{Finite element representation of cubes of cardiac tissue
    undergoing simple shear in the NS mode. The bottom of the cube is
    fixed and the top displacement is given. Left: homogeneous
    deformation with a constant shear angle. Right: finite element
    solution on a $6 \times 6 \times 6$ mesh. The plot shows the
    value of the NS-component of the right Cauchy Green strain tensor
    $\mathbf{C}$.}
  \label{fig:mesh}
\end{figure}
\end{center}

Additionally, we model the case of a homogeneous deformation which
corresponds to a linear displacement with a constant shear angle
throughout the domain. Such a model can be represented by discretizing
the cubes with a single layer of linear finite elements: the resulting
displacement is completely determined by the prescribed boundary
conditions. Figure~\ref{fig:mesh} illustrates the two kinds of
deformations on cube meshes.

The discrete variational formulation of the Euler-Lagrange equations
is implemented using the FEniCS Project
software~\cite{alnaes2014unified,LoggMardalEtAl2011a} and
dolfin-adjoint~\cite{farrell2013automated}. From a FEniCS forward
model, dolfin-adjoint automatically generates the symbolic adjoint
system of equations and computes the functional
gradient~\eqref{eq:adj_grad} using the adjoint solution. The FEniCS
framework automatically generates and compiles efficient C++ code for
the assembly of the relevant linear systems from the symbolic
representations of both forward and adjoint equations, and solves the
nonlinear and linear systems using
e.g.~PETSc~\cite{PETScPackage}. With this setup, we observed that a
typical solution of the Euler-Lagrange
equation~\eqref{eq:varform_fully} takes $6$ Newton iterations.

\section{Numerical Results}
\label{sec:results}

\subsection{Verification}
Each of the finite element, adjoint, and optimization solvers have been
carefully verified, separately and combined, as follows:

(i) The finite element solver was verified by the method of manufactured
solutions \cite{salari2000code}. Following this method we chose an analytic expression
for the displacement and pressure fields
\begin{equation}
\begin{array}{l}
\disp = \left( t x^3, \ y \left( \frac{1}{3tx^2 + 1} -1 \right) , 0 \ \right) \\
p = 0.
\end{array}
\label{eq:mms_solution}
\end{equation}
Here $x,y$ refer to Cartesian coordinates
and $t$ is a scaling parameter which we set to $t = 0.2$. Using this
analytic expression we derived Dirichlet boundary
conditions over a unit cube, and a loading term
$f$ which satisfied a pointwise form of equation~\eqref{eq:weakform}
\begin{equation}
 \frac{\partial \psi(\rcg, \mvec)}{\partial {\bf F}} + p J {\bf F}^{-T} = f \quad \mbox{in} \Omega.
 \label{eq:anatest}
\end{equation}
Note that the chosen displacement field satisfies the
  incompressibility constraint $J -1 = 0$. We then computed finite
  element approximations to \eqref{eq:mms_solution} and observed the
  expected second-order convergence of the displacement gradient to
  the analytical displacement gradient~\cite{hood1974navier}.

(ii) We verified the
computation of stresses in the finite element model by prescribing a
homogeneous deformation and comparing the resulting numerically
integrated top face shear stress values to analytically computed
values. The analytic values were based on the calculations found
in~\cite[Section 5a]{Holzapfel2009} and the numerical values were
observed to match closely.

(iii) We confirmed the correctness of the
adjoint gradients by considering the linearization of the functional $\hat{I}(\mvec)$
around $\mvec$ with perturbation $\Delta \mvec$ and using Taylor's theorem: the expression
\begin{equation}
\hat{I}(\mvec) - \hat{I}(\mvec + \Delta \mvec) + \frac{D \hat{I}(\mvec)}{D \mvec} (\Delta \mvec) = O(\Delta \mvec^2)
\end{equation}
converged to $0$ at a rate of $2$ as $\Delta \mvec \longrightarrow 0$, which can only
be expected if $\frac{D \hat{I}(\mvec)}{D \mvec}$ is computed accurately.

\subsection{Parameter estimation with synthetic data}

Additionally, we verified the optimization solver by performing a
synthetic data test. In this test we chose a target set of
  material parameters, Table~\ref{tab:inv_crime}, 2nd line, and used
  them to compute synthetic integrated stress values for all 6 shear
  modes of the tissue experiment~\cite{dokos2002shear}. These
  synthetic stresses were then matched by an optimization starting
  from material parameter values $25 \%$ higher than the target.

We performed this test using our two models for deformation.
The first model assumed a homogeneous shear
angle through the material and
the second model was a finite element model
with a $1 \times 1 \times 1$ mesh.
Since the displacement field of the finite element model
was element-wise quadratic, it allowed for more
flexibility in the deformation field. The results of this synthetic data
test are presented in Table~\ref{tab:inv_crime} and show that
the optimization algorithm was able to closely match the target material parameters.

\begin{table}
\caption{Synthetic data test results. The first row (Initial) contains
  the material parameter values used to initialize the algorithm,
  while the second row (Target) contains the parameters that were used
  to generate the synthetic stresses. The rows marked 'Homogeneous'
  and 'Finite Element' contain optimized parameter values coming from
  homogeneous deformation and finite element models. These optimized
  values are matched perfectly by the optimized homogeneous model and
  very closely by the finite element model.}
  \begin{tabular}{lrrrrrrrrr}
  \toprule
   & $a$ & $b$ & $a_f$ & $b_f$ & $a_s$ & $b_s$ & $a_{fs}$ & $b_{fs}$ & $I$ \\
   & (kPa)&  & (kPa)& & (kPa)& & (kPa)& & (mN) \\
  \midrule
  Initial & 0.059 &  8.023 &  18.472 &  16.026 & 2.481 &  11.120 &  0.216 &  11.436 & \\
  Target (80\%) & 0.047 &  6.418 &  14.778 & 12.821 & 1.985 &  8.896 &  0.173 &  9.149 &\\
  Homogeneous            & 0.047 & 6.418 & 14.778 & 12.821 & 1.985 & 8.896 & 0.173  & 9.149  & 4.611 $\times 10^{-8}$\\
  Finite Element         & 0.047 & 6.406 & 14.778 & 12.821 & 1.983 & 8.938 & 0.173  & 9.155 & 0.00082\\
  \bottomrule
  \end{tabular}
  \label{tab:inv_crime}
\end{table}

\subsection{Parameter estimation with experimental stress data}
In the following, we present the results of fitting the
Holzapfel-Ogden strain energy law~\eqref{eq:hao} using the objective
function~\eqref{eq:gaussfit} and a SLSQP optimizer with
bound constraints. The SLSQP algorithm makes use of the gradient of the
objective functional which we obtain using the adjoint gradient
formula~\eqref{eq:adj_grad}.

As the numerical solution of the nonlinear Euler-Lagrange
equation~\eqref{eq:varform_fully} easily fails to converge when a
material parameter becomes too small, we set a lower bound of $1.0
\times 10^{-2}$ on the components of $\mvec$ while optimizing finite
element models. This bound was not necessary for the homogeneous
deformation models as no Euler-Lagrange equation is solved. All
optimizations were carried out until the optimizer was unable to
further reduce the objective functional or an absolute tolerance of
$1.0\times 10^{-6}$ in the 2-norm of the functional gradient was reached.

\subsubsection{Material parameter estimation using \emph{a priori} knowledge}
\label{sub:a_priori}

The material parameters of the Holzapfel-Ogden model have previously
been estimated using a homogeneous deformation model (Table 1, 2nd row in~\cite{Holzapfel2009}).
We first used these values as the initial
values for optimization of our homogeneous model targeting the
transversely isotropic and orthotropic data sets. The optimized
results are listed in Table~\ref{tab:opt_results} with the label
Homogeneous.

\begin{figure}[htbp]
  \includegraphics[width=1.0\textwidth]{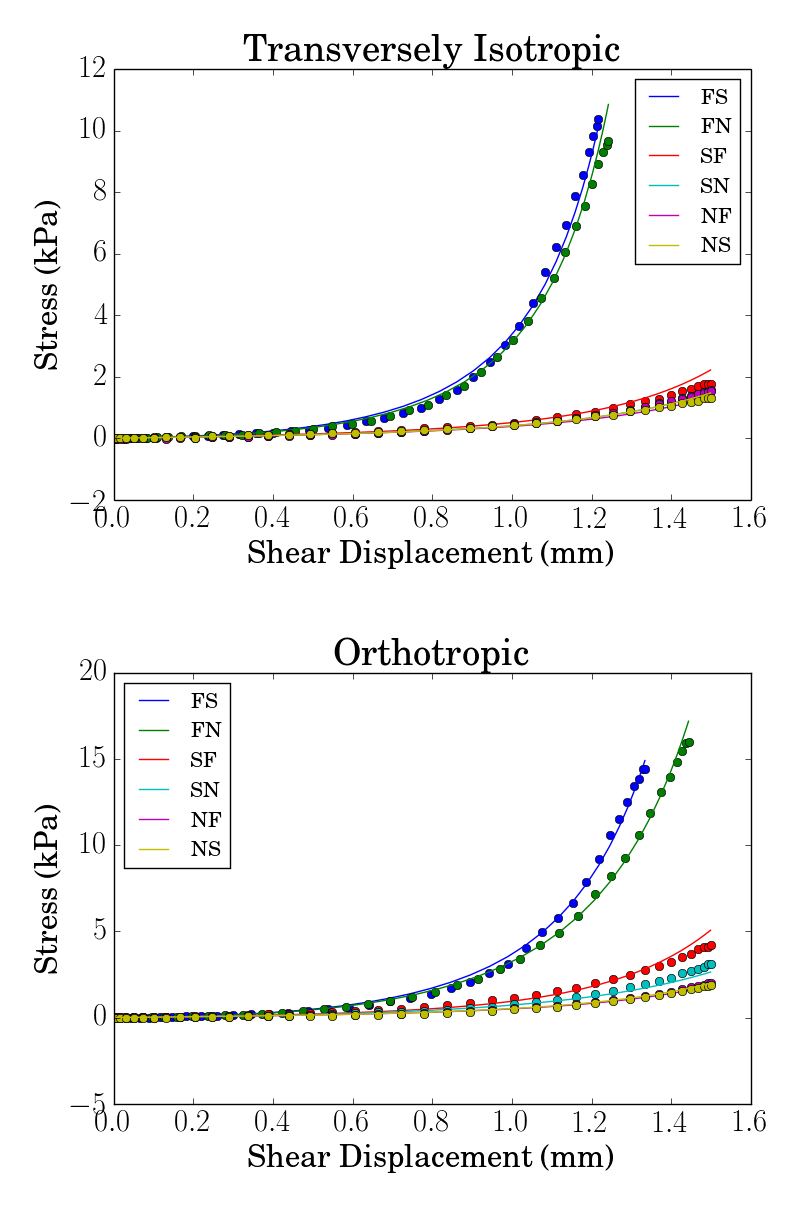}
  \caption{Comparison of optimized model stress-strain curves with
    experimental data.  The dots are interpolated experimental data at
    Gauss points, the solid lines show the output of the finite
    element models with $N = 8$ elements per edge of the cube.}
  \label{fig:exper_results}
\end{figure}

We next consider finite element models that allow for heterogeneous shear displacements.
Beginning with a $1 \times 1 \times
1$ cube and the optimal material parameters from the homogeneous model
as initial values, we computed optimal values for the $1 \times 1
\times 1$ case. This procedure was repeated for $N \times N \times N$
cubes with $N = 2, 4, 6, 8$, using the results of the previous
optimization as the initial condition for the next
case. The resulting parameter values are presented in
Table~\ref{tab:opt_results}, and the corresponding optimal
stress-strain curves are shown in Figure~\ref{fig:exper_results}.

We note that going from $N = 8$ to $N = 10$ using both the transversely
isotropic and the orthotropic
data does not change the material parameters rounded to two 2 significant digits,
and therefore consider our finite element models
to be sufficiently refined at this resolution. We also note that the fit values, $I$,
decreased with mesh refinement up to about 2 digits accuracy. We expect this decrease
since increased mesh refinement gives more
flexibility in the deformation field of the finite element model.

\begin{center}
\begin{table}
  \begin{tabular}{lrrrrrrrrrrr}
    \toprule
  & $a$ & $b$ & $a_f$ & $b_f$ & $a_s$ & $b_s$ & $a_{fs}$ & $b_{fs}$ & $I$ & Ev. & Grad \\
  & (kPa)&  & (kPa)& & (kPa)& & (kPa)& & (mN) & & Ev.\\
    \midrule
  \textbf{Transversely Isotropic} &        &         &         &         &       &       &          &  & & &\\
  Homogeneous             & 0.544 & 6.869 & 23.220 & 39.029  & 0.0001   & 0.172  & 0.248 & 5.310 & 3.291 & 41 & 21 \\
  N = 1                   & 0.593 & 6.841  & 23.209  & 38.826  & 0.010   & 0.010 & 0.243 & 9.531    & 3.173 & 44 & 37 \\
  N = 2                   & 0.732 & 6.818  & 22.110  & 39.946  & 0.010   & 0.010 & 0.183 & 13.614 & 3.010 & 24 & 18 \\
  N = 4                   & 0.807 & 6.737  & 21.349  & 40.468  & 0.010   & 0.010 & 0.122 & 17.936 & 2.819 & 25 & 18 \\
  N = 6                   & 0.794 & 6.859  & 21.212 & 40.537  & 0.010   & 0.010 & 0.129 & 17.462 & 2.802 & 22 & 15 \\
  N = 8                   & 0.784 & 6.973 & 21.149  & 40.584  & 0.010   & 0.010 & 0.145 & 16.401   & 2.815 & 21 & 14 \\
  N = 10           	  & 0.778 & 7.048 & 21.112 & 40.585 & 0.010 & 0.010 & 0.150 & 16.036 & 2.819 & 24 & 17 \\
 \midrule
  \textbf{Orthotropic} &        &         &         &         &         &    &          &          & & &\\
  Homogeneous  & 0.556  & 7.940  & 33.366 & 14.224 &  2.804 & 0.0001 & 0.588  & 8.216  & 6.804 & 31 & 20 \\
  N = 1        & 0.766  & 6.857  & 31.640 & 15.210 &  2.069 & 0.010  & 0.352  & 15.243 & 5.880 & 29 & 19 \\
  N = 2        & 1.040  & 6.557  & 29.375 & 15.979 &  1.742 & 0.010  & 0.118  & 23.296 & 4.565 & 39 & 24 \\
  N = 4        & 0.979  & 7.364 & 28.882& 15.813 &  2.058 & 0.010  & 0.107  & 24.039 & 3.952 & 28 & 16 \\
  N = 6  & 0.961 & 7.495 & 28.762 & 15.783 & 2.088 & 0.010 & 0.114 & 23.549 & 3.899 & 21 & 13\\
  N = 8  & 0.962 & 7.510 & 28.649 & 15.806 & 2.044 & 0.010 & 0.122 & 23.027 & 3.899 & 20 & 11 \\
  N = 10 & 0.959 & 7.542 & 28.565 & 15.813 & 2.017 & 0.010 & 0.123 & 22.750 & 3.981 & 25 & 12 \\
  \bottomrule
  \end{tabular}
  \caption{Material parameters fitted to the orthotropic and
    transversely isotropic datasets for the Homogeneous and $N \times
    N \times N$ finite element models.  $I$ refers to the value of the
    objective functional. The number of functional evaluations (Ev.) and
    functional gradient evaluations (Grad Ev.)
    are given in the two rightmost columns.}
  \label{tab:opt_results}
\end{table}
\end{center}

\subsubsection{Material parameter estimation using multi-start optimization}

In this section, we present the results of using the multi-start
method to estimate the optimal material parameters, rather than
relying on a good initial guess. For the calculation of random initial
guesses we set $P_{max} = 40$, cf.~Section~\ref{sec:multistart}.  This
value is close to the largest material parameter found in
Table~\ref{tab:opt_results}. Note that this choice gives a
conservative set of initial parameters for the optimization algorithm
(low initial values) which in turn enhances the robustness of the
procedure. We also set $60$ as an upper bound for each material
parameter value during the optimization. Without this upper bound we
observed that many optimizations crashed or converged to suboptimal
local minima.

In each multi-start experiment, $30$ random starting points were
  used. The mesh fineness was set to the level of $N = 8$, which was
  sufficient to give converged material parameter sets when using a
  priori knowledge in Section~\ref{sub:a_priori}.  In
  Table~\ref{tab:randopt_results} we present the best fitting results
  of the multi-start experiments and note that they are very close to
  those obtained with a priori knowledge in
  Table~\ref{tab:opt_results}.

\begin{table}
\caption{Results of fitting material parameters to the transversely
  isotropic and orthotropic data sets using the multi-start method. The
  rows labeled 'Multistart Best Fit' correspond to the optimizations with the
  lowest misfit value $I$. The rows labeled '$N
  = 8$' are copied from Table~\ref{tab:opt_results} for reference.}
  \begin{tabular}{lrrrrrrrrr}
    \toprule
  & $a$ & $b$ & $a_f$ & $b_f$ & $a_s$ & $b_s$ & $a_{fs}$ & $b_{fs}$ & I\\
  & (kPa)&  & (kPa)& & (kPa)& & (kPa)& & (mN) \\
    \midrule
   \textbf{Transversely Isotropic} & & & & & & & & & \\
    \midrule
   \emph{N = 8} & \emph{0.784} & \emph{6.973} & \emph{21.149}  & \emph{40.584}  & \emph{0.010} & \emph{0.010} & \emph{0.145} & \emph{16.401} & \emph{2.815} \\
     Multistart Best Fit & 0.795 & 6.855 & 21.207 & 40.545 & 0.010 & 0.010 & 0.130 & 17.446 & 2.802 \\
   \noalign{\smallskip}\hline \noalign{\smallskip}
   \textbf{Orthotropic} & & & & & & & & & \\
   \hline\noalign{\smallskip}
 \emph{N = 8}  & \emph{0.962} & \emph{7.510} & \emph{28.649} & \emph{15.806} & \emph{2.044} & \emph{0.010} & \emph{0.122} & \emph{23.027} & \emph{3.899} \\
   Multistart Best Fit & 0.964 & 7.510 & 28.654 & 15.791 & 2.051 & 0.010 & 0.118 & 23.230 & 3.959 \\
   \midrule
  \end{tabular}
  \label{tab:randopt_results}
\end{table}

\subsubsection{Objective functional values for alternative material parameters}
Several other studies~\cite{Holzapfel2009, goktepe2011computational,
  wang2013structure} have used of the Dokos et al. 2002 shear
data~\cite{dokos2002shear} to calibrate the Holzapfel and Ogden strain
energy~\eqref{eq:hao}. These studies used homogenized deformation
models for the optimization. In Table~\ref{tab:compare_paramestimates}
we list the computed objective functional value of parameter sets
originating from previous studies using the orthotropic dataset and
finite element model $(N = 8)$. The results indicate that our
parameter set fits these data better than the previously computed
ones.

We also note that our finite element parameter set with finite element model has a better fit value than the
homogeneous parameter set with the homogeneous model. Indeed we expect
the finite element fit to be at least as good as the homogeneous fit, as
the finite element model allows for greater flexibility
in the the deformation field, above and beyond
that of the homogeneous model.

\begin{table}
 \caption{Holzapfel-Ogden law parameter estimates from this and
   previous studies.  $I_{fem}$ indicates the value of the fit function
   \eqref{eq:gaussfit} with model stresses from a finite element model ($N =8$), and
   $I_{hom}$ the value of the same fit function but with model stresses computed
   with a homogeneous deformation model.
   The material parameters of the last two rows
   originate from homogeneous and finite element model fits respectively in Table~\ref{tab:opt_results}.
   Note that objective functional ($I$-) values for parameter sets from other studies are
   obtained using the orthotropic data used in this study (experimental data),
   and not the data used in the studies the parameter sets originate from (digitized data).}
 \begin{tabular}{lrrrrrrrrrr}
   \toprule
   Source                      & $a$ & $b$   & $a_f$ & $b_f$   & $a_s$ & $b_s$ & $a_{fs}$ & $b_{fs}$ & $I_{hom}$ & $I_{fem}$\\
                               & (kPa)&  & (kPa)& & (kPa)& & (kPa)& & (mN) & (mN)\\
   \midrule
   Holzapfel et al 2009        & 0.059  & 8.023 & 18.472 & 16.026 & 2.481  & 11.120 & 0.216  & 11.436 & 36.143 & 36.825\\
   Goektepe et al 2011         & 0.496  & 7.209 & 15.193 & 20.417 & 3.283  & 11.176 & 0.662  & 9.466  & 28.583 & 29.480 \\
   Wang et al 2013             & 0.2362 & 0.810 & 20.037 & 14.154 & 3.7245 & 5.1645 & 0.4108 & 11.300 & 33.271 &  34.195 \\
   Current (hom)              & 0.556 & 7.940 & 33.366 & 14.224 & 2.804 & 0.0001 & 0.587 & 8.216 & \textbf{6.804} & 9.653 \\
   Current (fem)              & 0.962 & 7.510 & 28.649 & 15.806 & 2.044 & 0.010 & 0.122 & 23.027 & 41.622 & \textbf{3.899} \\
   \bottomrule
 \end{tabular}
 \label{tab:compare_paramestimates}
\end{table}

\subsection{Computational efficiency of the adjoint-based functional gradient}
\label{subsec:adj_efficiency}

\begin{figure}
  \includegraphics[width=1.0\textwidth]{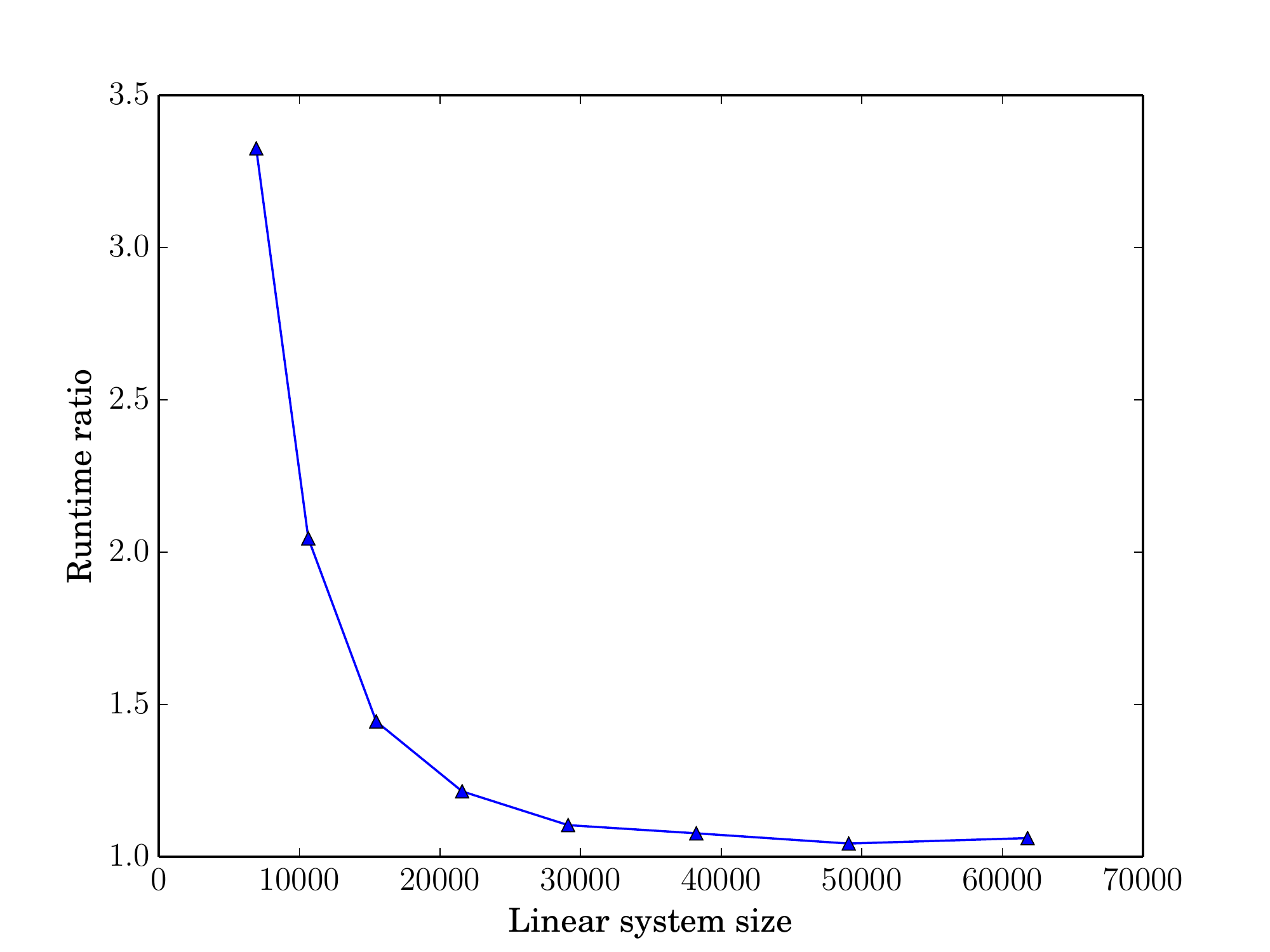}
  \caption{Gradient efficiency: ratio of gradient evaluation runtime
    over single Newton iteration runtime for increasing linear system
    sizes.}
  \label{fig:timing_exper_results}
\end{figure}

Adjoint solver efficiency may be measured by comparing the runtime of
the adjoint and forward solves. Here, we examine the overall gradient
efficiency in a similar manner. We consider the evaluation of the
gradient of the objective functional~\eqref{eq:gaussfit}, though in a
reduced case with only a single shear mode included in the sum and a
reduced forward solve consisting of a single nonlinear solver
iteration. In this case, the forward and adjoint models each consist
of a single linear solve in addition to a number of residual
evaluations. For larger linear system sizes, the runtime of a linear
solve is expected to dominate the runtime of assembly, and thus these
forward and adjoint models are of roughly the same computational
expense.

For this reduced case, we evaluated the adjoint-based gradient for a
range of linear system sizes. For each system size, we calculated the
gradient runtime ratio; that is, the runtime used by the evaluation of
the gradient divided by the runtime of the forward solve. The
resulting ratios are plotted in
Figure~\ref{fig:timing_exper_results}. The curve indicates that the
gradient run-time ratio gets close to the theoretically optimal value
of $1$ as we increase the system size.


\section{Discussion}
\label{sec:discussion}

\subsection{Choice of shearing experiment datasets}

Of the six shearing experiment datasets,
cf.~Figure~\ref{fig:shear_exprdata}, we have used two for parameter
estimation. One of the reasons for this choice is an incompatibility
of most of the datasets with assumptions made in the design of the
strain energy functional~\eqref{eq:hao}. In particular, the strain
energy~\eqref{eq:hao} dictates an ordering of the shear mode
stiffnesses in the case of a homogeneous shear displacement. We
  can see this by adapting the analysis that leads to equations (5.23)
  -- (5.28) of~\cite{Holzapfel2009}. In this analysis a parameter
  $\gamma$ is introduced to represent the amount of simple shear
  displacement present in a homogeneous deformation. For example for
  the FS mode
\begin{equation}
 {\bf F} = \begin{bmatrix}
1 \ & \gamma \ & 0 \ \\
0 \ & 1 \ & 0 \ \\
0 \ & 0 \ & 1 \ \end{bmatrix}.
\end{equation}
Using this deformation gradient, and the respective deformation gradients of the other modes,
the shear component of the Cauchy stress $\sigma$ in the shearing direction can be calculated for
each mode. If we consider the same invariants as in~\eqref{eq:hao}, that is $I_1, I_{4f}, I_{4s}, I_{8fs}$,
and use the notation $\psi_i = \frac{\partial \psi}{\partial I_{i}}$, we arrive at the following
equations for shear stress as a function of shear displacement
\begin{equation}
 \begin{aligned}
  \mbox{(FS):} \quad \sigma_{FS} &= 2(\psi_1 + \psi_{4f})\gamma + \psi_{8fs}, \\
  \mbox{(FN):} \quad \sigma_{FN} &= 2(\psi_1 + \psi_{4f})\gamma, \\
  \mbox{(SF):} \quad \sigma_{SF} &= 2(\psi_1 + \psi_{4s})\gamma + \psi_{8fs}, \\
  \mbox{(SN):} \quad \sigma_{SN} &= 2(\psi_1 + \psi_{4s})\gamma,\\
  \mbox{(NF):} \quad \sigma_{NF} &= 2\psi_1\gamma,\\
  \mbox{(NS):} \quad \sigma_{NS} &= 2\psi_1\gamma.\\
 \end{aligned}
 \label{eq:stressorderings}
\end{equation}
For further details regarding the derivation of these equations we refer the reader to \cite{Holzapfel2009}.
The simple shear stresses~\eqref{eq:stressorderings}
reveal two assumptions built into the design of~\eqref{eq:hao}, namely for homogeneous simple shear deformations
\begin{equation}
\begin{aligned}
 \sigma_{FS} \geq \sigma_{FN} \geq \sigma_{NF},\\
 \sigma_{SF} \geq \sigma_{SN} \geq \sigma_{NF}.\\
 \end{aligned}
 \label{eq:model_assumptions}
\end{equation}

Out of the six datasets, only one is consistent with these orderings,
namely the 6th one, which was used here under the label transversely
isotropic. In this dataset the stress-strain relationship is typical
of a transversely isotropic material with a stiffer fiber
direction. In several other cardiac mechanics simulation studies
\cite{krishnamurthy2013patient, gjerald2015patient,
  Finsberg:2015:Online}, the Holzapfel and Ogden energy
functional~\eqref{eq:hao} has been simplified to model transversely
isotropic behavior by removing the terms involving the invariants
$I_{4s}, I_{8fs}$. For such a simplified model one could use the
parameter estimates for $a,b, a_f, b_f$ that we obtained from the
Transversely Isotropic dataset.

However, the Holzapfel and Ogden model was originally proposed to
model orthotropic behavior. This motivates also targeting a dataset
displaying fully orthotropic properties. In particular, dataset 2 in
Figure~\ref{fig:shear_exprdata} is such and compares well with Figure
6 of~\cite{dokos2002shear} and Figure 2 of~\cite{Holzapfel2009}. By
switching the $SF$ and $SN$ curves of Dataset 2 we were able to
reinterpret this data in a way that is consistent with the
interpretation in~\cite{Holzapfel2009}, and the shear stiffness
orderings~\eqref{eq:model_assumptions}.

\subsection{Discussion of optimal material parameter values}

We have obtained two sets of material parameters: one corresponding to
an orthotropic case and one corresponding to a transversely isotropic
case. We observe that for both sets of material parameters, the $b_s$
parameter essentially vanishes. For the Transversely Isotropic case,
both $a_s$ and $b_s$ essentially vanish, which is in excellent
agreement with the transversely isotropic stress-strain
pattern. Furthermore we note that the magnitude of both $a_s$ and
  $b_s$ parameters in the best fitting parameter sets presented in
  Table~\ref{tab:randopt_results} are very small. In light of the
  shear stress calculations \eqref{eq:stressorderings} we can see that
  the $a_s$ and $b_s$ parameters are related to the degree of extra
  stiffness in the sheet direction over the sheet normal
  direction. Indeed when we examine the shear data,
  Figure~\ref{fig:exper_results}, we can see that the $SN-SF$ curves
  are only slightly stiffer than the $NF-NS$ curves, which explains
  why the optimal values of $a_s$ and $b_s$ are so small.

Comparing the orthotropic material parameter values to the previously
published values in Table~\ref{tab:compare_paramestimates}, we observe
that the fit of our material parameters is significantly better, as
expected. By using a finite element model we have been able to relax
the homogeneous shearing angle assumption and more realistically model
the motion of the cubes in the shearing experiment. We note that our
material parameters differ from those previously published, and also
that there is a significant variability in the parameter values
previously reported. Some of this variability is most likely due
  to the differences in the selection of points during the
  digitization of [Figure 2 of~\cite{Holzapfel2009}], which was done
  in the studies whose material parameter sets we compare in
  Table~\ref{tab:compare_paramestimates}.  By using original data from
  the shearing experiment, we were able to remove the uncertainty due
  to digitization in our parameter estimates.  Finally we note that
  even after the SF-SN curves are swapped in Dataset 2 of
  Figure~\ref{fig:shear_exprdata}, there are still minor differences
  when compared to [Figure 7 of~\cite{Holzapfel2009}] and [Figure 3
    of~\cite{goktepe2011computational}] and [Figure 4
    of~\cite{wang2013structure}]. This also explains why our parameter
  sets differ from those calculated in the previous studies. 

\subsection{Computing functional gradients in cardiac mechanics}

Figure~\ref{fig:timing_exper_results} demonstrates that the
computational cost of the adjoint gradient computation is comparable
to that of a single iteration of the nonlinear solution algorithm
of~\eqref{eq:varform_fully} for larger system sizes. For smaller
system sizes, the cost of symbolic computation and the cost of
residual and Jacobian assembly contribute significantly yielding
higher ratios -- as expected. Wang et al.'s 2013 simulations of a
human left ventricle in diastole use system sizes of approximately
$100\, 000$ degrees of freedom~\cite{wang2013structure}. Given the
trend in Figure~\ref{fig:timing_exper_results}, we can expect that the
adjoint method and solver implemented in this work will continue to be
efficient at this scale and beyond.


Comparatively, assuming the use of Newton's method for the solution of
nonlinear systems, the evaluation of a finite difference gradient
requires a linear system assembly and solve for each Newton iteration,
and one nonlinear solve is required per component of the gradient.
Counting the 8 parameters in the Holzapfel-Ogden model~\eqref{eq:hao},
and assuming a typical solution of the Euler-Lagrange
equation~\eqref{eq:varform_fully} takes 6 Newton iterations, we can
expect the computational cost of finite difference gradient evaluation
to be circa 48 times greater than that of the adjoint method.

In the optimization results of Table~\ref{tab:opt_results}, we
observed iteration counts of up to $44$
for the optimization of $8$
parameters using our gradient-based method. This compares favorably
with the circa $7000$ iterations needed to estimate $9$ parameters
using a global method in~[Figure 5 of \cite{wong2015velocity}].

\subsection{Implications for organ-scale image-based parameter estimation with spatially resolved
material parameters}

Although we have tested our adjoint-based multi-start optimization
method on the 2002 shear data of Dokos et al~\cite{dokos2002shear}, we
believe our methods will provide the biggest advantage in the case of
optimizing cardiac model parameters in high spatial resolution at the
organ scale to MRI or echocardiographic image data. In this case the high spatial
resolution would allow for detailed modelling of regional differences
in tissue stiffness, which is for example present in patients with post-infarct fibrosis.

In such an application
a model parameter could be represented as a finite element function similarly to
the displacement or hydrostatic pressure fields (\disp, p). Doing this would increase
the number of components of the gradient $\frac{D \hat{I}}{D \mvec}$ by the number of
degrees of freedom needed to spatially represent the parameter of interest.
Using a finite difference or reduced order Kalman filter
approach in this case would require an additional evaluation of
the Euler-Lagrange equation~\eqref{eq:varform_fully}
for each degree of freedom introduced,
whereas the adjoint gradient formula~\eqref{eq:adj_grad}
only needs to be calculated once
regardless of the number of additional degrees of freedom. In the current
study the adjoint gradient is estimated to be
\[\mbox{(number of model parameters)} \times 6 = 48\]
times faster than finite differencing.
In the case of a spatially varying model parameter the speedup is potentially
a lot more significant.


When fitting material parameters to the Dokos experiment data, we
  were able to generate good initial guesses for the local
  optimization by progressively refining the mesh and using the
  optimal results from the previous coarser refinement level as an
  initial guess in the successive finer level. It would be more
  challenging to apply this technique using image based ventricular
  geometries, due to the problem of accurately representing the
  geometry with few elements. As an alternative we propose the
  multi-start approach, which we have shown here to be accurate and
  viable using the Dokos experiment data.

One issue that would arise in using the multi-start approach with
image based geometries would be the choice of the number of
multi-start points; using less points is more computationally
efficient, while using more is potentially more robust.  Possible
solutions are the use of optimal stopping
criteria~\cite{boender1987bayesian} or more sophisticated local-global
searches~\cite{tsai2003global, goldberg1999optimizing}.


\section{Conclusions}
\label{sec:conclusion}

In this work, we have presented a new application of efficient
gradient-based optimization methods in the context of estimating
cardiac hyperelastic material parameters from experimental data. In
particular, we have demonstrated how an adjoint solution can greatly
speed up the evaluation of functional gradients. These methods have
produced two new sets of material parameter values that yield
simulated stress-strain curves that fit closely to orthotropic and
transversely isotropic shear data. For future parameter estimation
studies using image based geometries and a local search algorithm,
multi-start or a similar method should be used in order to avoid
suboptimal minima.

\begin{acknowledgements}
The authors would like to thank Socrates Dokos, Holger Schmid and Ian
LeGrice, for making the experimental data available. Our work is
supported by The Research Council of Norway through a Centres of
Excellence grant to the Center for Biomedical Computing at Simula
Research Laboratory, project number 179578, and also through the
Center for Cardiological Innovation at Oslo University Hospital
project number 203489.  Aln{\ae}s has been supported by the Research
Council of Norway through grant number 209951. Computations were
performed on the Abel supercomputing cluster at the University of Oslo
via NOTUR project NN9316k.
\end{acknowledgements}

\bibliographystyle{spmpsci}      
\bibliography{bibliography}   
\end{document}